\documentclass[lettersize,journal]{IEEEtran}
\usepackage{amsmath,amsfonts}
\usepackage{algorithm}
\usepackage{array}
\usepackage[caption=false,font=normalsize,labelfont=sf,textfont=sf]{subfig}
\usepackage{textcomp}
\usepackage{stfloats}
\usepackage{url}
\usepackage{verbatim}
\usepackage{graphicx}
\usepackage{cite}
\usepackage{amssymb} 
\usepackage{xcolor}
\usepackage{algorithmicx}
\usepackage{algpseudocode}
\usepackage{tabularx}  
\usepackage{booktabs}  % For \toprule, \midrule, \bottomrule

\usepackage{hyperref} 
\hypersetup{
	colorlinks=true,       
	linkcolor=blue,         
	anchorcolor=black,     
	citecolor=blue,        
	urlcolor=blue,         
}
\begin{document}
	
	\title{Near-Field Wideband Channel Estimation for XL-MIMO Systems via Denoising Diffusion Model}
	\author{Qingxia Feng,~\IEEEmembership{Graduate Student Member, IEEE,} Yin Fang, Meng Hua,~\IEEEmembership{Senior Member, IEEE,} Cheng Zhang,~\IEEEmembership{Senior Member, IEEE,} Chunguo Li,~\IEEEmembership{Senior Member, IEEE,} Yongming Huang,~\IEEEmembership{Fellow, IEEE,} and Luxi Yang,~\IEEEmembership{Senior Member, IEEE}
	%\thanks{This work was supported by the National Key R\&D Program of China under Grant 2025ZD1302500, the Natural Science Foundation on Frontier Leading Technology Basic Research Project of Jiangsu under Grants BK20222001, the National Natural Science Foundation of China under Grant U1936201, and Grant 61971128. \emph{(Corresponding author: Luxi Yang.)}}
		
    %\thanks{Qingxia Feng, Yin Fang, Cheng Zhang, Chunguo Li, Yongming Huang and Luxi Yang are with the School of Information Science and Engineering, Southeast University, Nanjing 210096, China, and also with the Purple Mountain Laboratories, Nanjing 211111, China (e-mail: \{qxfeng, yinfang, zhangcheng\_seu,chunguoli, huangym, lxyang\}@seu.edu.cn). }
	%\thanks{Meng Hua is with the Department of Electrical and Electronic Engineering, Imperial College London, SW7 2AZ London, U.K. (e-mail:m.hua@imperial.ac.uk).}
		
		}

\markboth{}%
{Shell \MakeLowercase{\textit{et al.}}: A Sample Article Using IEEEtran.cls for IEEE Journals}


\maketitle
\begin{abstract}
	Extremely large-scale multiple-input multiple-output (XL-MIMO) is a key enabling technology for sixth-generation (6G) communication systems. Nevertheless, the increase in array aperture and signal bandwidth brings new challenges to wideband channel estimation in XL-MIMO systems. Motivated by recent advances in deep generative modeling, we propose a diffusion model-based method for near-field wideband channel estimation in XL-MIMO systems. We first analyze the statistical correlation of wideband channel and show that near-field wideband channel exhibits both spatial non-stationarity and beam split effects. Based on these observations, the channel estimation problem is formulated as a Bayesian posterior inference task, in which a diffusion model is employed to learn the prior distribution of the channel. To further enhance the representation of complex spatial-frequency channel structures, we design a denoising network with a multi-scale attention mechanism. In particular, the network extracts multi-scale spatial-frequency features via parallel convolutional branches with different receptive fields, and combines feature attention and spatial attention modules to adaptively emphasize critical channel features. This design enables more accurate modeling of near-field wideband channel distributions and consequently improves channel estimation performance. Experimental results demonstrate that the proposed method exhibits superior robustness to existing baseline schemes for XL-MIMO wideband channel estimation under different experimental settings.
\end{abstract}

\begin{IEEEkeywords}
	XL-MIMO, wideband channel estimation, near-field, diffusion model.
\end{IEEEkeywords}

\section{Introduction}
\IEEEPARstart{W}{ith} the evolution of sixth-generation (6G) communication systems, extremely large-scale multiple-input multiple-output (XL-MIMO) has been widely recognized as a key enabling technology for enhancing the spectral efficiency, spatial multiplexing capability, and communication reliability of wireless networks \cite{10098681}. By equipping hundreds or even thousands of antennas at the base station (BS), XL-MIMO can significantly improve array gain and spatial resolution \cite{10858129}. Meanwhile, high-frequency bands such as millimeter-wave (mmWave) offer abundant bandwidth for high-data-rate transmission \cite{5783993,11153494}. The short wavelengths at high frequencies enable the integration of more antenna elements within a limited physical space, while the large beamforming gain of XL-MIMO helps mitigate the severe path loss in wideband communications. As a result, XL-MIMO wideband communication has become one of the most important research directions in the development of 6G systems.

In XL-MIMO systems, tasks such as beamforming and resource allocation rely heavily on accurate channel state information (CSI), making high-precision channel estimation essential \cite{10786241,10271123,10541333,10938996,9738442}. As the array size increases, the Rayleigh distance also grows, causing more user terminals to fall into the near-field region. Under such conditions, electromagnetic wave propagation exhibits spherical-wave characteristics, and the channel response depends on both the angle of arrival and the propagation distance. In addition, under wideband transmission, the array response varies across subcarriers with frequency and exhibits a significant beam split effect \cite{10791414,10486834,11176026,10857463}. Consequently, conventional channel models based on the far-field plane-wave assumption may be inadequate to accurately capture the actual propagation characteristics, thereby making wideband channel estimation in XL-MIMO systems more challenging.

Recently, near-field channel estimation has attracted significant attention \cite{11122492,Cui2022_Channel,r450,10373799,10780971,10634218}. Among conventional linear estimators, the least squares (LS) method is simple to implement and does not require prior statistical information. However, it is sensitive to noise and performs poorly in low signal-to-noise ratio (SNR) regimes. The linear minimum mean square error (LMMSE) estimator can achieve better performance by exploiting second-order channel statistics, but its effectiveness relies heavily on accurate prior information, such as the channel covariance, and its computational complexity is high due to the inversion of large-dimensional matrices. To address the limitations of linear estimators and the challenges posed by near-field propagation, recent studies have employed compressed sensing (CS) methods with a polar-domain dictionary to characterize channel sparsity in both the angular and distance domains \cite{11122492}. Using uniform sampling in the angular domain and non-uniform sampling in the distance domain, \cite{Cui2022_Channel} demonstrated the sparsity of near-field channels in the polar domain and proposed corresponding channel estimation methods based on simultaneous orthogonal matching pursuit (SOMP). Moreover, sparse Bayesian learning has been introduced to achieve high-precision sparse recovery of beamspace channels, thereby facilitating near-field channel estimation \cite{r450}.

Besides near-field propagation, XL-MIMO systems also exhibit significant spatial non-stationarity due to their extra-large array aperture. Unlike conventional massive MIMO systems, where each multipath component is typically observable by the entire array, different propagation paths are often visible only to a subset of antennas, resulting in a non-uniform distribution of channel statistics across the array. To address spatial non-stationarity in XL-MIMO systems, existing studies have proposed a variety of channel estimation methods. Specifically, subarray partitioning and group time block code (GTBC) were employed in \cite{10373799} for non-stationary channel estimation. In addition, hierarchical sparse models and Markov chain-based priors were introduced in \cite{10780971} to jointly characterize spatial non-stationarity in the angular and spatial domains. Based on these modeling approaches, a two-stage orthogonal matching pursuit (TL-OMP) algorithm was further developed for non-stationary channel estimation in \cite{10634218}. 

However, most existing studies have not fully addressed the challenges caused by frequency-dependent array responses in near-field wideband communications. In wideband XL-MIMO systems, the steering vectors vary with frequency across subcarriers, leading to inconsistent sparse support structures \cite{10500431,9444239,10713259,11029409,10709906}. To address this effect, the beam split pattern detection (BSPD) method exploits the linear variation of sparse support with subcarrier frequency and establishes the corresponding beam-split pattern for wideband channel recovery \cite{9444239}. Based on the linear structure of the polar sparse support set across subcarriers, the bilinear pattern detection (BPD) method has also been proposed for accurate recovery of near-field wideband XL-MIMO channels \cite{10713259}. In addition, \cite{11029409} developed a two-stage wideband channel estimation method based on polar sparsity, which performs power-based sub-interval coarse estimation and window-detection-based fine estimation. A frequency-dependent polar dictionary with adaptive angle and distance sampling has been proposed for wideband channel covariance estimation, along with adaptive matching pursuit and parameter refinement techniques \cite{10709906}. Nevertheless, these CS-based methods still rely heavily on hand-crafted priors and predefined dictionaries, which often lead to substantial storage and computational overhead in near-field wideband scenarios.

To address the complex nonlinear characteristics of near-field channel propagation, various deep learning-based methods have been proposed \cite{10143629,10967069,8752012,mao2025neft}. These methods can reduce the reliance on explicit sparsity assumptions to some extent. For example, a model-driven approach based on a fixed point network combines orthogonal approximate message passing (OAMP) with nonlinear estimation aided by a neural network to improve channel estimation performance \cite{10143629}. A deep unfolding network based on a frequency dependent polar dictionary has also been developed by formulating channel estimation over multiple subcarriers as a multiple measurement vector recovery problem with common sparse support and exploiting joint sparsity across subcarriers through threshold shrinkage \cite{10967069}. In addition, deep convolutional neural network (CNN) has been employed to capture local spatial features across adjacent antennas and subcarriers, thereby improving estimation accuracy \cite{8752012}. However, most existing methods based on feedforward neural networks formulate channel estimation as a direct regression task. Given pilot observations or partial measurements, the corresponding conditional channel distribution is often highly complex and difficult to characterize using simple parametric models. Consequently, these methods may lack sufficient expressive capability to accurately represent complex channel distributions.

Given these challenges, diffusion models, as a rapidly evolving class of deep generative models, can progressively capture the intrinsic structure of complex data distributions by learning data-driven implicit priors \cite{meng2022diffusion,song2023pseudoinverse,song2021denoising,pmlr-v258-fesl25a}. This capability enables the characterization of channels with rich underlying structures without relying on a specific model, while reducing reliance on predefined sparsity assumptions. Compared with methods relying on hand-crafted sparse priors or simple deterministic mappings, diffusion models provide a promising framework for modeling complex near-field wideband XL-MIMO channel distributions and achieving high-accuracy channel recovery. Motivated by these advantages, this paper investigates an algorithm based on a diffusion model for near-field wideband channel estimation. The main contributions are summarized as follows:
\begin{itemize}
	\item We first analyze the statistical correlation of near-field wideband channel. The results show that, under spherical-wave propagation, near-field wideband channel exhibits both spatial non-stationarity and beam split effects, leading to highly non-uniform channel statistics over the spatial and frequency domains. Specifically, the antenna correlation depends not only on the element spacing but also on the absolute antenna position. This means that inter-antenna correlations vary with the antenna index, which reveals the spatial non-stationarity of the channel. Moreover, the subcarrier correlation is also affected by the antenna position, implying that the frequency-domain statistical structure varies across space and further reflecting the coupling between beam split and spatial location.
	
	\item To address the difficulty of accurately characterizing the complex and non-uniform distribution of near-field wideband channel using hand-crafted priors, we propose a diffusion model-based framework for near-field wideband channel estimation. From a Bayesian posterior inference perspective, channel estimation is formulated as a posterior recovery problem, where a diffusion model is used to learn the generative prior of the channel distribution. By incorporating the measurement likelihood, the proposed framework enables iterative channel recovery under posterior score guidance, thereby providing a data-driven solution for near-field wideband channel estimation.
	
	\item To further enhance the ability of the diffusion model to capture the spatial-frequency structure of near-field wideband channel, we develop a denoising network with a multi-scale attention mechanism. Specifically, parallel convolutional branches with different receptive fields are employed to extract multi-scale spatial-frequency features, while feature attention and spatial attention modules are introduced to adaptively emphasize critical channel structures. This design strengthens the representation capability of the diffusion model and thereby improves channel recovery accuracy.
	
	\item Numerical experiments demonstrate that the proposed diffusion model based wideband channel estimation algorithm achieves remarkable robustness under different bandwidths, transmission distances, and noise conditions.
\end{itemize}

The remainder of this paper is organized as follows. Section II presents the system and channel model. Section III analyzes the statistical correlation of near-field wideband channel. Section IV details the proposed generative diffusion model-based algorithm for XL-MIMO channel estimation. Section V presents the simulation results. Finally, Section VI concludes the paper.

\section{System Model and Problem Formulation}
Consider the uplink transmission in a wideband XL-MIMO system. The BS is equipped with a uniform linear array (ULA) consisting of $N$ antenna elements and communicates with single-antenna user equipment over $M$ subcarriers using orthogonal frequency division multiplexing (OFDM). The spacing between adjacent antenna elements is set to $d=\lambda_c/2$, where $\lambda_c=c/f_c$ denotes the wavelength corresponding to the carrier frequency, with $c$ and $f_c$ representing the speed of light and the carrier frequency, respectively.

\begin{figure}[!t]
	\centering
	\includegraphics[width=2.5in]{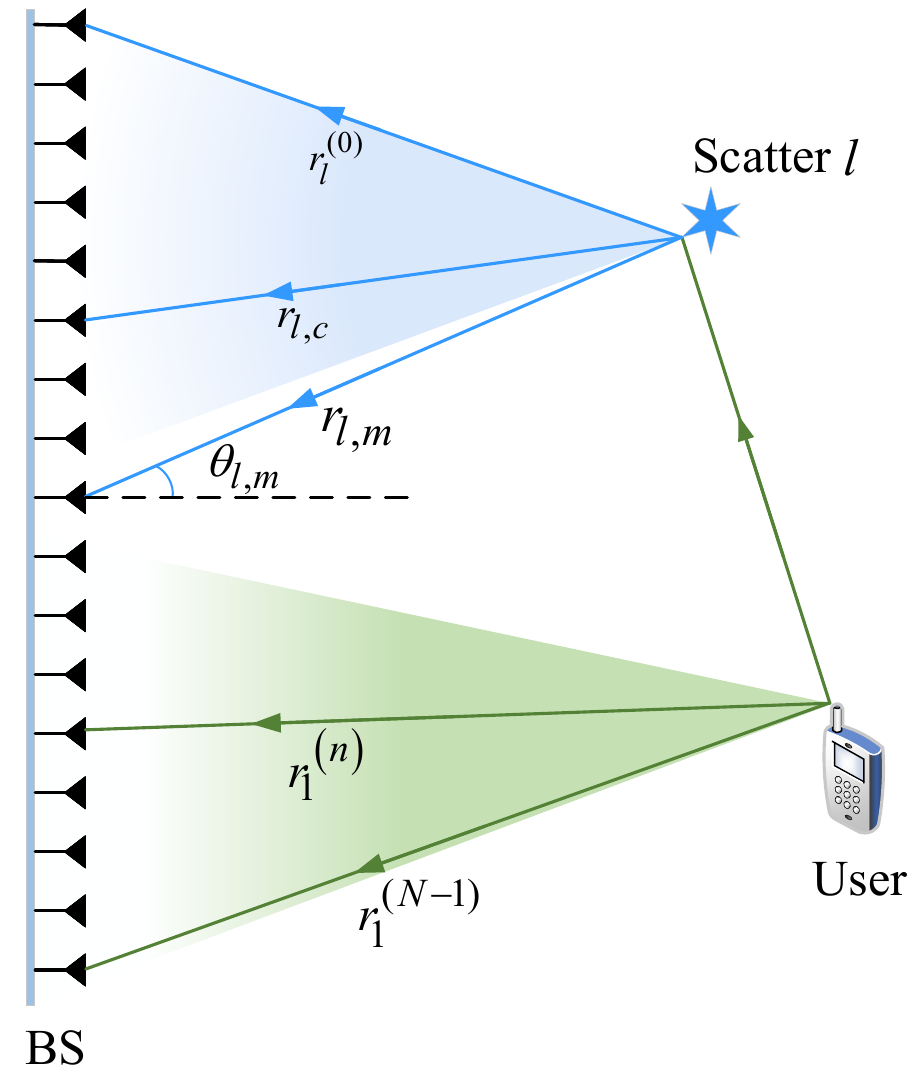}
	\caption{System model of the near-field wideband channel.}
	\label{fig:3_system_model}
\end{figure}

\subsection{Near-field Channel Model}
To characterize near-field propagation in XL-MIMO systems, a geometric multipath channel model is adopted \cite{11029409,10486834}, as shown in Fig.~\ref{fig:3_system_model}. Let $r_l$ denote the distance from the user or scatterer to the antenna origin along the $l$th propagation path. Then, the near-field channel vector $\mathbf{h}[m]\in\mathbb{C}^{N\times 1}$ on the $m$th subcarrier is expressed as \cite{10709906,10500431}
\begin{equation}\label{eq:3_channel}
	\mathbf{h}[m] = \sqrt{\frac{N}{L}} \sum_{l=1}^{L} \alpha_{l} \, e^{-j\frac{2\pi f_m}{c} r_l} \, \mathbf{a}(\theta_l, r_l, f_m),
\end{equation}
where $L\ll N$ is the number of distinguishable multipath components, and $\alpha_l$ denotes the complex path gain. The frequency of the $m$th subcarrier is
\begin{equation}
	f_m = f_c + \frac{B}{M}\left(m - 1 - \frac{M-1}{2}\right),
\end{equation}
where $B$ and $f_c$ denote the system bandwidth and carrier frequency, respectively, and the corresponding wavelength is $\lambda_m=c/f_m$. Moreover, $\theta_l=\sin\phi_l$ is the spatial angle parameter associated with the physical incidence angle $\phi_l$ of the $l$th path. Let $(\theta_{l,m}, r_{l,m})$ denote the angle and distance corresponding to $(\theta_l,r_l)$ at subcarrier frequency $f_m$.

In XL-MIMO systems, the enlarged array size and aperture $D$ often place user equipment in the near-field region, where the propagation distance is smaller than the Rayleigh distance $2D^2/\lambda_c$. In this regime, conventional far-field channel models based on the plane-wave assumption are no longer sufficient, and a spherical-wave propagation model is required.

Accordingly, the normalized array steering vector $\mathbf{a}(\theta_l,r_l,f_m)\in\mathbb{C}^{N\times1}$ is given by
\begin{equation}
	\mathbf{a}(\theta_l, r_l, f_m) =
	\frac{1}{\sqrt{N}}
	\left[
	e^{j\varphi_{l,m}^{(0)}},
	e^{j\varphi_{l,m}^{(1)}},
	\ldots,
	e^{j\varphi_{l,m}^{(N-1)}}
	\right]^T ,
\end{equation}
where the phase of the $n$th antenna element is
\begin{equation}
	\varphi_{l,m}^{(n)} = \frac{2\pi}{\lambda_m}\left(r_l - r_l^{(n)}\right).
\end{equation}
Here, $r_l^{(n)}$ denotes the propagation distance from the scatterer to the $n$th BS antenna along the $l$th path, which is given by	$r_l^{(n)} = r_l + n^2 d^2 \xi_l - n d \theta_l$ with $\xi_l = \frac{1-\theta_l^2}{2r_l}$\cite{Cui2022_Channel}. 

Consequently, the near-field channel model differs fundamentally from the conventional far-field model. Unlike traditional massive MIMO systems, where channel characterization mainly relies on angular-domain parameters, XL-MIMO channels depend on both angle and propagation distance, thus requiring joint angle-distance modeling. As the array aperture $D$ increases significantly in XL-MIMO systems, the corresponding Rayleigh distance can reach tens or even hundreds of meters. For example, in a system with $256$ antennas operating at $60$ GHz, the Rayleigh distance is approximately $162.56\,\mathrm{m}$, implying that many user devices may lie in the near-field region. Therefore, near-field effects must be taken into account in both system modeling and channel estimation.

Moreover, in wideband systems, the array steering vector varies with frequency. While the narrowband assumption approximates $f_m \approx f_c$ and hence $\mathbf{a}(\theta_l,r_l,f_m) \approx \mathbf{a}(\theta_l,r_l,f_c)$, this approximation is no longer valid in wideband scenarios. The frequency variation across subcarriers introduces coupling among $\theta_l$, $r_l$, and $f_m$, leading to the beam split effect \cite{10541333,10786241,10271123}. This effect further increases the difficulty of near-field wideband channel estimation, and neglecting it causes model mismatch and performance degradation \cite{10500431}. Therefore, the dependence of $\mathbf{a}(\theta_l,r_l,f_m)$ on $f_m$ is explicitly considered in the adopted channel model.

\subsection{Signal Model and Problem Formulation}
Based on \eqref{eq:3_channel}, the received signal on the $m$th subcarrier is given by \cite{11029409,11176026,10857463}
\begin{equation}
	\mathbf{y}[m] = \sqrt{\frac{N}{L}}
	\sum_{l=1}^{L}
	\alpha_{l}
	e^{-j\frac{2\pi f_m}{c} r_l}
	\mathbf{a}(\theta_l,r_l,f_m)
	s[m]
	+
	\mathbf{n}[m],
\end{equation}
where $s[m]$ is the pilot symbol transmitted on the $m$th subcarrier, and $\mathbf{n}[m]\in\mathbb{C}^{N\times1}$ denotes the additive white Gaussian noise (AWGN) vector following $\mathcal{CN}(\mathbf{0},\sigma_n^2\mathbf{I}_N)$, with $\sigma_n^2$ being the noise power.

The received signals over all $M$ subcarriers can be stacked into the matrix form
\begin{equation}
	\mathbf{Y}=\mathbf{H}\mathbf{S}+\mathbf{N},
\end{equation}
where $\mathbf{S}=\operatorname{diag}(s[1],s[2],\cdots,s[M])\in\mathbb{C}^{M\times M}$ denotes the pilot diagonal matrix with $|s[m]|^2=P_t$ for all $m$, $\mathbf{Y}=[\mathbf{y}[1],\mathbf{y}[2],\cdots,\mathbf{y}[M]]\in\mathbb{C}^{N\times M}$ is the received signal matrix, $\mathbf{H}=[\mathbf{h}[1],\mathbf{h}[2],\cdots,\mathbf{h}[M]]\in\mathbb{C}^{N\times M}$ is the channel matrix, and $\mathbf{N}=[\mathbf{n}[1],\mathbf{n}[2],\cdots,\mathbf{n}[M]]\in\mathbb{C}^{N\times M}$ is the noise matrix. By vectorizing the above equation, we obtain
\begin{equation}\label{eq:signal}
	\mathbf{y}=\mathbf{B}\mathbf{h}+\mathbf{n},
\end{equation}
where $\mathbf{y}=\mathrm{vec}(\mathbf{Y})$, $\mathbf{h}=\mathrm{vec}(\mathbf{H})$, $\mathbf{n}=\mathrm{vec}(\mathbf{N})$, and $\mathbf{B}=\mathbf{S}^T\otimes\mathbf{I}_N$.

Accordingly, given the known pilot matrix $\mathbf{S}$, the channel estimation problem is to recover the unknown channel vector $\mathbf{h}$ from the observation $\mathbf{y}$. A conventional approach is the LS estimator, which is formulated as
\begin{equation}\label{LS}
	\hat{\mathbf{h}}_{LS}=\mathrm{vec}(\hat{\mathbf{H}}_{LS})=\arg\min_{\mathbf{h}}\left\|\mathbf{y}-\mathbf{Bh}\right\|^2.
\end{equation}
with the closed-form solution
\begin{equation}
	\hat{\mathbf{h}}_{\mathrm{LS}}=\frac{1}{\sqrt{P_t}}(\mathbf{B}^H\mathbf{B})^{-1}\mathbf{B}^H\mathbf{y}.
\end{equation}
By further exploiting the second-order statistics of the channel, the LMMSE estimator can be expressed as
\begin{equation}
	\hat{\mathbf{h}}_{\mathrm{LMMSE}}=\mathbf{R}_{\mathbf{h}\mathbf{h}}\left(\mathbf{R}_{\mathbf{h}\mathbf{h}}+\frac{\sigma_n^2}{P_t}\mathbf{I}_{NM}\right)^{-1}\hat{\mathbf{h}}_{\mathrm{LS}}.
\end{equation}
Although the LMMSE estimator is mean-square-error optimal under perfectly known channel second-order statistics, its practical applicability is limited by the difficulty of obtaining accurate channel correlation information in realistic propagation environments, as well as the high computational complexity incurred by large-dimensional matrix inversion.

\section{Analysis of Near-Field Wideband Channel Characteristics}
This section analyzes the statistical characteristics of near-field wideband channel and reveals their non-uniform correlation structures across both the antenna and frequency domains. Based on this analysis, the limitations of conventional methods built upon handcrafted priors are identified, which in turn justifies the adoption of generative diffusion models for learning channel distribution.

\subsection{Correlation between Antennas}
From the near-field channel model in \eqref{eq:3_channel}, the channel response between the user and the $n$th BS antenna on the $m$th subcarrier is given by 
\begin{equation}\label{eq:channel_element}
	[\mathbf{h}[m]]_n=\sqrt{\frac{1}{L}}\sum_{l=1}^{L}\alpha_{l}e^{-j\frac{2\pi f_{m}}{c}r_{l}}\cdot e^{j\frac{2\pi f_{m}}{c}\left(nd\theta_{l}-n^{2}d^{2}\xi_{l}\right)}.
\end{equation}

Since near-field wideband channels depend on both spatial angle and propagation distance, their statistical correlation can be analyzed from these two dimensions jointly. In practice, each resolvable path is composed of multiple unresolvable subpaths, whose angles and propagation distances are distributed around their mean values \cite{1146527}. Let $\phi$ denote the offset from the mean physical angle $\bar{\phi}_0$, such that $\phi_0=\bar{\phi}_0-\phi$ and $\theta_0=\sin(\bar{\phi}_0-\phi)$. Likewise, let $\psi$ denote the offset from the mean distance $\bar{r}_0$, such that $r_0=\bar{r}_0+\psi$. Then, on subcarrier $m$, the spatial correlation between the $n$th and $(n+\Delta_n)$th antennas is given by
\begin{equation}
	\begin{aligned}
		R_a(n,f_m)=\mathbb{E}_{\phi,\psi}\left[[\mathbf{h}[m]]_n\cdot[\mathbf{h}^H[m]]_{n+\Delta_n}\right].
	\end{aligned}
\end{equation}
Substituting equation \eqref{eq:channel_element} into the above expression yields
\begin{equation}\label{eq:R_a}
	\begin{aligned}
		R_a(n,f_m)
		=&\frac{\sigma_{\alpha}^{2}}{L}\sum_{l=1}^{L}\mathbb{E}_{\phi,\psi}\left[e^{j\frac{2\pi f_{m}}{c}\left[(2n\Delta_n+\Delta_n^2)d^2\xi_l-\Delta_nd\theta_l\right]}\right]\\
		=&\sigma_{\alpha}^{2}\iint_{\phi,\psi}\mathcal{P}_r(\psi)\mathcal{P}_\theta(\phi)\cdot\\&
		e^{j\frac{2\pi f_{m}}{c}\left[(2n\Delta_n+\Delta_n^2)d^2\xi_0-\Delta_nd\theta_0\right]}\mathrm{d}\phi\mathrm{d}\psi.
	\end{aligned}
\end{equation}
%To analyze the inter-antenna correlation, we first consider the phase term in \eqref{eq:R_a}. Due to the high spatial resolution of XL-MIMO, the angular spread within each resolvable path is small \cite{592600}. The trigonometric terms can be approximated using a first-order Taylor expansion as follows\begin{equation}	\theta_0=\sin(\bar{\phi}_0-\phi)\approx\sin\bar{\phi}_0-\phi\cos\bar{\phi}_0=\bar{\theta}_0-\phi\cos\bar{\phi}_0,\end{equation}\begin{equation}	1-\theta_0^2=\cos^2(\bar{\phi}_0-\phi)\approx\cos^2\bar{\phi}_0+\phi\sin2\bar{\phi}_0,\end{equation}where $\bar{\theta}_0=\sin\bar{\phi}_0$. Likewise, the distance term is approximated as $\frac{1}{r_0}=\frac{1}{\bar{r}_0+\psi}\approx\frac{1}{\bar{r}_0}\left(1-\frac{\psi}{\bar{r}_0}\right)$. Substituting the above approximations into $\xi_0=\frac{1-\theta_0^2}{2r_0}$ yields\begin{equation}	\xi_{0}=\frac{1-\theta_{0}^{2}}{2r_{0}}\approx\bar{\xi}_{0}+\frac{\phi\sin2\bar{\phi}_{0}}{2\bar{r}_{0}}-\frac{\bar{\xi}_{0}}{\bar{r}_{0}}\psi,\end{equation}where $\bar{\xi}_0=\frac{1-\bar{\theta}_0^2}{2\bar{r}_0}=\frac{\cos^2\bar{\phi}_0}{2\bar{r}_0}$.

To analyze the inter-antenna correlation, we first examine the phase term in \eqref{eq:R_a}, which can be decomposed into a deterministic component, an angular-dependent component, and a distance-dependent component, as follows
\begin{equation}
	\begin{aligned}
		&\frac{2\pi f_{m}}{c}\left[(2n\Delta_n+\Delta_n^2)d^2\xi_0-\Delta_nd\theta_0\right]\\
		=&\frac{2\pi f_m\Delta_nd\cos\bar{\phi}_0}{c}\phi+\frac{2\pi f_m(2n\Delta_n+\Delta_n^2)d^2\sin2\bar{\phi}_0}{2c\bar{r}_0}\phi\\&
		-\frac{2\pi f_m\Delta_nd\bar{\theta}_0}{c}+\frac{2\pi f_m(2n\Delta_n+\Delta_n^2)d^2\bar{\xi}_0}{c}
		\\
		&-\frac{2\pi f_m(2n\Delta_n+\Delta_n^2)d^2\bar{\xi}_0}{c\bar{r}_0}\psi,
	\end{aligned}
\end{equation}
where $\bar{\theta}_0=\sin\bar{\phi}_0$ and $\bar{\xi}_0=\frac{1-\bar{\theta}_0^2}{2\bar{r}_0}=\frac{\cos^2\bar{\phi}_0}{2\bar{r}_0}$. 

For notational convenience, we define $\Omega_a=\frac{f_m\Delta_nd\cos\bar{\phi}_0}{c}+\frac{f_m(2n\Delta_n+\Delta_n^2)d^2\sin2\bar{\phi}_0}{2c\bar{r}_0}$. Then, \eqref{eq:R_a} can be expressed as
\begin{equation}
	\begin{aligned}
		R_a(n,f_m)=\sigma_\alpha^2\cdot e^{j[-\frac{2\pi f_m\Delta_nd\bar{\theta}_0}{c}+\frac{2\pi f_m(2n\Delta_n+\Delta_n^2)d^2\bar{\xi}_0}{c}]}\cdot\mathcal{I}_\theta^a\cdot\mathcal{I}_r^a,
	\end{aligned}
\end{equation}
where
\begin{equation}
	\mathcal{I}_{\theta}^{a}=\int_{-\pi}^{\pi}\mathcal{P}_{\theta}(\phi)e^{j2\pi\Omega_{a}\phi}\mathrm{d}\phi,
\end{equation}
and
\begin{equation}
	\mathcal{I}_r^a=\int_0^\infty\mathcal{P}_r(\psi)e^{-j\frac{2\pi f_m(2n\Delta_n+\Delta_n^2)d^2\bar{\xi}_0}{c\bar{r}_0}\psi}\mathrm{d}\psi.
\end{equation}

We next evaluate $\mathcal{I}_\theta^a$ and $\mathcal{I}_r^a$. The $\mathcal{P}_{\theta}(\phi)$ can be interpreted as the power angle spectrum \cite{Forenza2007_Simplified}, which is expressed as
\begin{equation}\label{eq:PAS}
	\mathcal{P}_\theta(\phi)=\frac{\beta}{\sqrt{2}\sigma_\phi}\exp\left(-\frac{\sqrt{2}|\phi|}{\sigma_\phi}\right),\quad\phi\in[-\pi,\pi),
\end{equation}
where $\sigma_{\phi}$ is the standard deviation and $\beta=\frac{1}{1-e^{-\sqrt{2}\pi/\sigma_{\phi}}}$ is the normalization factor. This leads to
\begin{equation}
	\begin{aligned}
	\mathcal{I}_\theta^a=&\frac{\sqrt{2}\sigma_\phi\beta}{2+\sigma_\phi^2(2\pi\Omega_a)^2}(e^{-\frac{\sqrt{2}\pi}{\sigma_\phi}}[-\frac{\sqrt{2}}{\sigma_\phi}\cos(2\pi^2\Omega_a)\\&
	+2\pi\Omega_a\sin(2\pi^2\Omega_a)]+\frac{\sqrt{2}}{\sigma_\phi}).
	\end{aligned}
\end{equation}

The $\mathcal{P}_r(\psi)$ is modeled by the power delay profile \cite{753726,1622098}, given by
\begin{equation}\label{eq:PDP}
	\mathcal{P}_r(\psi)=K_r\exp\left(-\frac{\psi}{\sigma_\psi}\right),\quad\psi>0,
\end{equation}
where $K_{r}=1/\sigma_{\psi}$ is a normalization factor and $\sigma_{\psi}$ denotes the standard deviation. Then,
\begin{equation}
	\mathcal{I}_r^a=\frac{1}{1+j\sigma_\psi\frac{2\pi f_{m}(2n\Delta_{n}+\Delta_{n}^{2})d^{2}\bar{\xi}_{0}}{c\bar{r}_{0}}},
\end{equation}
and
\begin{equation}
	|\mathcal{I}_r^a|=\frac{1}{\sqrt{1+\left[\frac{2\pi f_{m}(2n\Delta_{n}+\Delta_{n}^{2})d^{2}\bar{\xi}_{0}\sigma_{\psi}}{c\bar{r}_{0}}\right]^{2}}}.
\end{equation}
\begin{figure}[!t]
	\centering
	\includegraphics[width=3.3in]{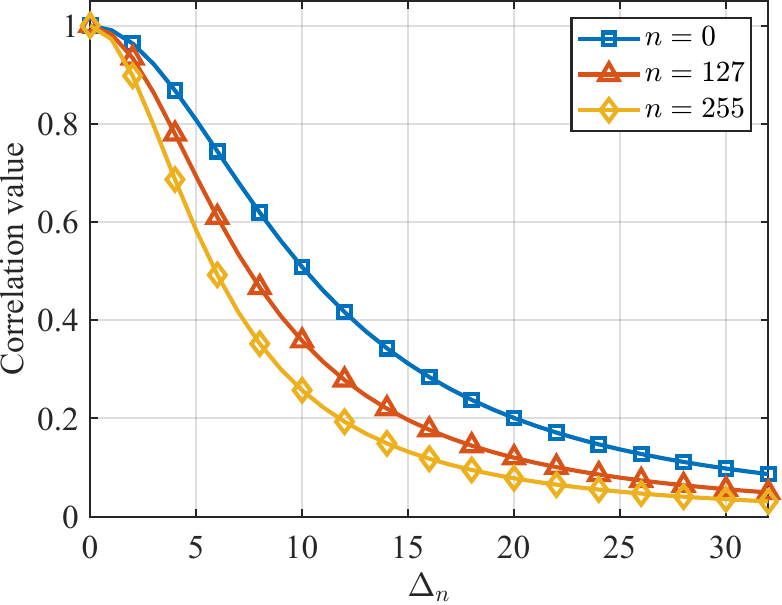}
	\caption{The spatial correlation value.}
	\label{fig:3_a_Correlation}
\end{figure}

Therefore, we have
\begin{equation}\label{eq:antenna_correlation}
	\begin{aligned}
		|R_{a}(n,f_{m})|=\sigma_{\alpha}^{2}\cdot|\mathcal{I}_\theta^a|\cdot\frac{1}{\sqrt{1+\left[\frac{2\pi f_{m}(2n\Delta_{n}+\Delta_{n}^{2})d^{2}\bar{\xi}_{0}\sigma_{\psi}}{c\bar{r}_{0}}\right]^{2}}},
	\end{aligned}
\end{equation}
which shows that the inter-antenna correlation in near-field wideband channel is jointly determined by the $|\mathcal{I}_\theta^a|$ and $|\mathcal{I}_r^a|$, both of which explicitly depend on the absolute antenna index $n$. As a result, even for a fixed antenna spacing $\Delta_n$, the correlation generally varies across different array positions, i.e., $R_a(n_1,f_m)\neq R_a(n_2,f_m), \quad n_1 \neq n_2$. As shown in Fig.~\ref{fig:3_a_Correlation}, the antenna correlation is no longer determined solely by the relative antenna spacing, but also varies with the absolute position of the antennas in the array.

This observation reveals the spatial non-stationarity of near-field channel, where the effective correlation range changes across array elements. By contrast, in conventional far-field models, the channel correlation depends only on the antenna spacing and is therefore spatially stationary. Such position-dependent statistical characteristics introduce additional challenges for channel estimation in near-field wideband XL-MIMO systems.

\subsection{Correlation between Subcarriers}
Based on the near-field channel model in \eqref{eq:3_channel}, the correlation between the $m$th and $(m+\Delta_m)$th subcarriers at the $n$th antenna is defined as
\begin{equation}\label{eq:R_s1}
	R_s(\Delta_m,n)=\mathbb{E}_{\phi,\psi}\left[[\mathbf{h}[m]]_n\cdot[\mathbf{h}^H[m+\Delta_m]]_n\right].
\end{equation}
Let $f_s=B/M$ denote the subcarrier spacing, such that $f_{m+\Delta_m}=f_m+\Delta_m f_s$. Substituting \eqref{eq:channel_element} into the \eqref{eq:R_s1} yields
\begin{equation}
	\begin{aligned}\label{eq:R_s}
		R_s(\Delta_m,n)
		=&\frac{\sigma_{\alpha}^{2}}{L}\sum_{l=1}^{L}\mathbb{E}_{\phi,\psi}\left[e^{-j\frac{2\pi\Delta_mf_s}{c}(nd\theta_l-n^2d^2\xi_l-r_l)}\right]\\
		=&\sigma_{\alpha}^{2}\iint_{\phi,\psi}\mathcal{P}_r(\psi)\mathcal{P}_\theta(\phi)\cdot \\&e^{-j\frac{2\pi\Delta_mf_s}{c}(nd\theta_0-n^2d^2\xi_0-r_0)}\mathrm{d}\phi\mathrm{d}\psi.
	\end{aligned}
\end{equation}

To analyze the inter-subcarrier correlation, we first examine the phase term in \eqref{eq:R_s}, which can be rewritten as
\begin{equation}
	\begin{aligned}
		-&\frac{2\pi\Delta_mf_s}{c}(nd\theta_0-n^2d^2\xi_0-r_0)\\=&\frac{2\pi\Delta_mf_s}{c}\left(\bar{r}_0-nd\bar{\theta}_0+n^2d^2\bar{\xi}_0\right)\\&
			+\frac{2\pi\Delta_{m}f_{s}}{c}\left(nd\cos\bar{\phi}_{0}+\frac{n^{2}d^{2}\sin2\bar{\phi}_{0}}{2\bar{r}_{0}}\right)\phi\\&
			+\frac{2\pi\Delta_mf_s}{c}\left(1-\frac{n^2d^2\bar{\xi}_0}{\bar{r}_0}\right)\psi.
	\end{aligned}
\end{equation}
 
From the above equation, it can be seen that the phase term in \eqref{eq:R_s} can be decomposed into a deterministic component determined by the mean parameters, together with two stochastic components associated with the angular offset $\phi$ and the distance offset $\psi$, respectively. Accordingly, the inter-subcarrier correlation can be expressed as
\begin{equation}
	\begin{aligned}
		R_s(\Delta_m,n)&=\sigma_\alpha^2\cdot e^{j\frac{2\pi\Delta_mf_s}{c}\left(\bar{r}_0-nd\bar{\theta}_0+n^2d^2\bar{\xi}_0\right)}\cdot\mathcal{I}_\theta^s\cdot\mathcal{I}_r^s.
	\end{aligned}
\end{equation}

To further analyze $\mathcal{I}_\theta^s$ and $\mathcal{I}_r^s$, we define $\Omega_s=\frac{\Delta_mf_s}{c}\left(nd\cos\bar{\phi}_0+\frac{n^2d^2\sin2\bar{\phi}_0}{2\bar{r}_0}\right)$. $\mathcal{I}_\theta^s$ and $\mathcal{I}_r^s$ are expressed as follows
\begin{equation}
	\mathcal{I}_{\theta}^{s}=\int_{-\pi}^{\pi}\mathcal{P}_{\theta}(\phi)e^{j2\pi\Omega_{s}\phi}\mathrm{d}\phi,
\end{equation}
and
\begin{equation}
	\begin{aligned}
		\mathcal{I}_r^s&=\int_0^\infty\mathcal{P}_r(\psi)e^{j\frac{2\pi\Delta_mf_s}{c}\left(1-\frac{n^2d^2\bar{\xi}_0}{\bar{r}_0}\right)\psi}\mathrm{d}\psi.
	\end{aligned}
\end{equation}

Therefore, based on \eqref{eq:PAS}, $\mathcal{I}_\theta^s$ can be expressed as
\begin{equation}
	\begin{aligned}
	\mathcal{I}_\theta^s=&\frac{\sqrt{2}\sigma_\phi\beta}{2+\sigma_\phi^2(2\pi\Omega_s)^2}(e^{-\frac{\sqrt{2}\pi}{\sigma_\phi}}[-\frac{\sqrt{2}}{\sigma_\phi}\cos(2\pi^2\Omega_s)
	\\&+2\pi\Omega_s\sin(2\pi^2\Omega_s)]+\frac{\sqrt{2}}{\sigma_\phi}).
	\end{aligned}
\end{equation}

Similarly, based on \eqref{eq:PDP}, $\mathcal{I}_r^s$ is given by
\begin{equation}
	\begin{aligned}
		\mathcal{I}_r^s&=\frac{1}{1-j\frac{2\pi\Delta_mf_s}{c}\left(1-\frac{n^2d^2\bar{\xi}_0}{\bar{r}_0}\right)\sigma_\psi},
	\end{aligned}
\end{equation}
and
\begin{equation}
	|\mathcal{I}_r^s|=\frac{1}{\sqrt{1+\left[\frac{2\pi\Delta_mf_s\sigma_\psi}{c}\left(1-\frac{n^2d^2\bar{\xi}_0}{\bar{r}_0}\right)\right]^2}}.
\end{equation}

Therefore,
\begin{equation}\label{eq:f_correlation}
	|R_s(\Delta_m,n)|=\sigma_\alpha^2\cdot|\mathcal{I}_\theta^s|\cdot\frac{1}{\sqrt{1+\left[\frac{2\pi\Delta_mf_s\sigma_\psi}{c}\left(1-\frac{n^2d^2\bar{\xi}_0}{\bar{r}_0}\right)\right]^2}}.
\end{equation}

\begin{figure}[!t]
	\centering
	\includegraphics[width=3.3in]{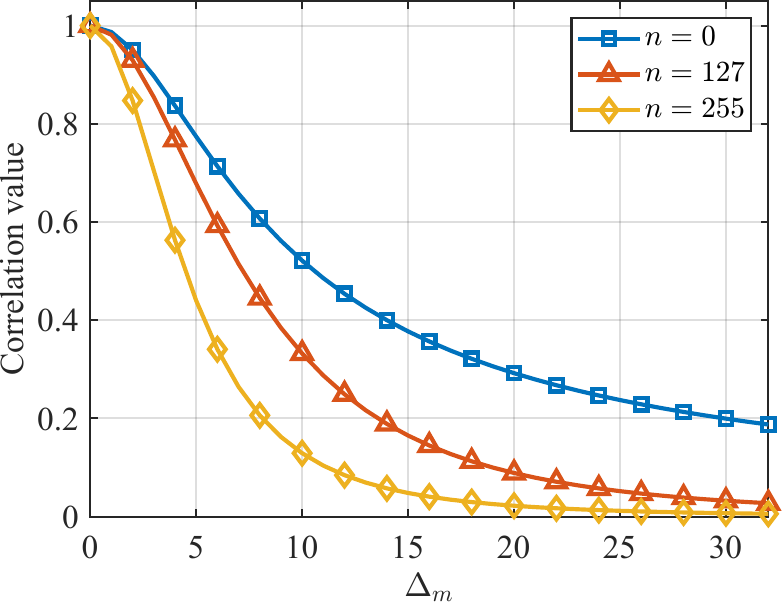}
	\caption{The frequency correlation value.}
	\label{fig:3_f_Correlation}
\end{figure}

Expression \eqref{eq:f_correlation} shows that the inter-subcarrier correlation in near-field wideband channel depends jointly on the subcarrier spacing $\Delta_m$ and the antenna index $n$. This space-frequency coupling originates from the beam split effect under wideband transmission. In particular, the angular-domain coefficient $\Omega_s$ contains the term $\frac{\Delta_m f_s n d \cos\bar{\phi}_0}{c}$, which captures a spatial shift that depends on the antenna position and is induced by frequency variation across subcarriers. Consequently, even for a fixed subcarrier spacing $\Delta_m$, the correlation varies across array positions, i.e., $|R_s(\Delta_m,n_1)| \neq |R_s(\Delta_m,n_2)|,\quad n_1 \neq n_2$, as illustrated in Fig.~\ref{fig:3_f_Correlation}. This implies that the frequency-domain channel statistics are spatially non-uniform across the array, which introduces additional challenges for near-field wideband channel estimation.

From the perspective of channel correlation, near-field wideband XL-MIMO channels exhibit intricate structural properties induced by spherical-wave propagation, spatial non-stationarity, and beam split. As a result, conventional methods developed under plane-wave and stationarity assumptions often suffer from model mismatch, while handcrafted priors, such as the $l_{0}$ sparsity adopted in the OMP algorithm \cite{Cui2022_Channel}, are insufficient to characterize the highly non-uniform distribution of such channels. Furthermore, the channel distribution is generally too complex to be accurately captured by simple parametric models, which limits the effectiveness of direct regression using feedforward neural networks. By contrast, generative diffusion model provides a flexible approach to learning intrinsic channel structures without relying on specific model assumptions, thereby offering a promising solution for near-field wideband channel estimation. Motivated by these observations, the next section presents the proposed diffusion model-based channel estimation method.

\section{Diffusion Model-Based Channel Estimation Method from a Bayesian Perspective}
This section develops a diffusion model-based framework for near-field wideband channel estimation. The main idea is to learn the prior distribution of wireless channel from offline training samples using a diffusion model, and then combine the learned generative prior with the measurement likelihood during online inference to recover the target channel. To better capture the non-uniform spatial-frequency characteristics of near-field wideband XL-MIMO channel, a denoising network with a multi-scale attention mechanism is further designed to enhance the representation capability of the diffusion model.

\subsection{A Diffusion Model-Based Framework for Near-Field Wideband Channel Estimation}

For the observation model in \eqref{eq:signal}, channel estimation aims to recover the unknown channel $\mathbf{h}$ from the observation $\mathbf{y}$. From a Bayesian perspective, this problem can be cast as posterior inference, i.e.,
\begin{equation}
	p(\mathbf{h}|\mathbf{y})\propto p_0(\mathbf{h})\,p(\mathbf{y}|\mathbf{h}),
\end{equation}
where $p_0(\mathbf{h})$ denotes the channel prior distribution and $p(\mathbf{y}\mid\mathbf{h})$ is the likelihood function. Conventional methods often rely on manually specified priors, such as the $l_0$ sparsity adopted in orthogonal matching pursuit (OMP) \cite{Cui2022_Channel}. However, near-field wideband channel usually exhibits complex and spatially non-uniform structures, which are difficult to characterize accurately using simple hand-crafted priors. To address this issue, pre-trained diffusion model is employed as generative priors and combined with posterior sampling for channel recovery \cite{meng2022diffusion}.

\subsubsection{Training Process}
The proposed diffusion model learns the data distribution by constructing a Markov chain that progressively perturbs the data with Gaussian noise. To account for the correlation between the real and imaginary parts, the complex-valued channel is reformulated as a real-valued representation. Given a prior distribution $\mathbf{h}_0\sim p_0(\cdot)$, the forward diffusion process is defined as
\begin{equation}
	p(\mathbf{h}_t|\mathbf{h}_{t-1})
	=
	\mathcal{N}\!\left(
	\mathbf{h}_t;
	\sqrt{1-\beta_t}\mathbf{h}_{t-1},
	\beta_t\mathbf{I}
	\right),\quad t\in[T],
\end{equation}
where $\beta_t$ is the predefined noise schedule satisfying $0<\beta_1<\cdots<\beta_T<1$. Define $\gamma_t=1-\beta_t$ and $\bar{\gamma}_t=\prod_{i=1}^{t}\gamma_i$. Then, the noisy sample at time step $t$ can be written as
\begin{equation}\label{eq:laten_re}
	\mathbf{h}_t
	=
	\sqrt{\bar{\gamma}_t}\mathbf{h}_0
	+
	\sqrt{1-\bar{\gamma}_t}\boldsymbol{\epsilon}_t,
\end{equation}
where $\boldsymbol{\epsilon}_t\sim\mathcal{N}(\mathbf{0},\mathbf{I})$. For $t=T$, since $\bar{\gamma}_T\approx 0$, the sample $\mathbf{h}_T$ approximately follows the standard Gaussian distribution $\mathcal{N}(\mathbf{0},\mathbf{I})$.

The proposed diffusion model learn the reverse denoising process corresponding to the forward noising process, thereby enabling the generation of samples from the original data distribution starting from noisy inputs. This reverse denoising process is characterized by the conditional distribution $p(\mathbf{h}_{t-1}|\mathbf{h}_t)$, which is generally intractable because the true data distribution $p_0(\cdot)$ is unknown. To approximate this reverse process, a denoising network $\boldsymbol{\epsilon}_{\boldsymbol{\theta}}(\mathbf{h}_t,t)$ parameterized by $\boldsymbol{\theta}$ is employed to predict the injected noise. The training objective is defined as
\begin{equation}\label{eq:loss_re}
	\mathcal{L}(\boldsymbol{\theta})
	=
	\mathbb{E}_{\mathbf{h}_0,\boldsymbol{\epsilon}_t,t}
	\left[
	\left\|
	\boldsymbol{\epsilon}_t
	-
	\boldsymbol{\epsilon}_{\boldsymbol{\theta}}
	\!\left(
	\sqrt{\bar{\gamma}_t}\mathbf{h}_0
	+
	\sqrt{1-\bar{\gamma}_t}\boldsymbol{\epsilon}_t,
	t
	\right)
	\right\|_2^2
	\right],
\end{equation}
where $t\sim\mathcal{U}([T])$. This objective trains the network to accurately estimate the noise component at different diffusion steps.

In each training iteration, a clean channel sample $\mathbf{h}_0$ is drawn from the dataset, a Gaussian noise sample $\boldsymbol{\epsilon}\sim\mathcal{N}(\mathbf{0},\mathbf{I})$ is generated, and the corresponding noisy sample $\mathbf{h}_t$ is constructed according to \eqref{eq:laten_re}, where the diffusion step $t$ is uniformly sampled from $[T]$. The denoising network $\boldsymbol{\epsilon}_{\boldsymbol{\theta}}$ takes $\mathbf{h}_t$ and $t$ as inputs and predicts the additive noise, and its parameters are optimized by minimizing \eqref{eq:loss_re}. Notably, this training procedure relies only on offline channel samples and does not involve observation data or specific noise statistics. Therefore, the trained diffusion model effectively captures the generative prior of the channel distribution. The architecture of the denoising network is detailed in Section \ref{sec:network}.

\subsubsection{Inference Process}
During channel recovery, the generative prior learned by the pre-trained diffusion model is incorporated into a Bayesian posterior inference framework to solve the inverse problem in \eqref{eq:signal}. For a given observation $\mathbf{y}$, the goal is to infer the posterior score of the noisy variable $\mathbf{h}_t$ at each diffusion step, i.e., $\nabla_{\mathbf{h}_t}\log p_t(\mathbf{h}_t|\mathbf{y})$.

According to Bayes' theorem, the posterior score can be decomposed as
\begin{equation}\label{eq:baye}
	\nabla_{\mathbf{h}_t}\log p_t(\mathbf{h}_t|\mathbf{y})
	=
	\nabla_{\mathbf{h}_t}\log p_t(\mathbf{h}_t)
	+
	\nabla_{\mathbf{h}_t}\log p_t(\mathbf{y}|\mathbf{h}_t).
\end{equation}
Here, $p_t(\mathbf{h}_t)$ and $\nabla_{\mathbf{h}_t}\log p_t(\mathbf{h}_t)$ represent the distribution of the noise-perturbed data at time step $t$ and its corresponding prior score, respectively. $p_t(\mathbf{y}|\mathbf{h}_t)$ and $\nabla_{\mathbf{h}_t}\log p_t(\mathbf{y}|\mathbf{h}_t)$ represent the noise-perturbed likelihood function and its corresponding likelihood score, respectively.

From the forward diffusion process in \eqref{eq:laten_re}, the conditional distribution is given by $p_{t}(\mathbf{h}_{t}|\mathbf{h}_{0})=\mathcal{N}(\sqrt{\bar{\gamma}_{t}}\mathbf{h}_{0},(1-\bar{\gamma}_{t})\mathbf{I})$, and its score function can be expressed as $\nabla_{\mathbf{h}_{t}}\log p_{t}(\mathbf{h}_{t}|\mathbf{h}_{0})=-\frac{\mathbf{h}_{t}-\sqrt{\bar{\gamma}_{t}}\mathbf{h}_{0}}{1-\bar{\gamma}_{t}}=-\frac{\boldsymbol{\epsilon}_t}{\sqrt{1-\bar{\gamma}_{t}}}$. The pre-trained denoising network $\boldsymbol{\epsilon}_{\boldsymbol{\theta}}(\mathbf{h}_t,t)$, which is trained to predict the noise $\boldsymbol{\epsilon}_t$, can approximate the prior score \cite{song2020score}.
\begin{equation}\label{eq:priori}
	\nabla_{\mathbf{h}_t}\log p_t(\mathbf{h}_t)
	\approx
	-\frac{1}{\sqrt{1-\bar{\gamma}_t}}
	\boldsymbol{\epsilon}_{\boldsymbol{\theta}}(\mathbf{h}_t,t).
\end{equation}

Since the likelihood score $\nabla_{\mathbf{h}_t}\log p_t(\mathbf{y}|\mathbf{h}_t)$ is generally intractable to compute exactly when $t>0$ \cite{song2023pseudoinverse}, the non-informative prior approximation method in \cite{meng2022diffusion} is adopted to transform the posterior distribution $p(\mathbf{h}_{0}|\mathbf{h}_{t})$. With a non-informative prior on $\mathbf{h}_0$, Bayes’ rule gives $p(\mathbf{h}_0|\mathbf{h}_t)\propto p(\mathbf{h}_t|\mathbf{h}_0)\cdotp(\mathbf{h}_0)\propto p(\mathbf{h}_t|\mathbf{h}_0)$. The posterior distribution is approximately given by $p(\mathbf{h}_0|\mathbf{h}_t)\approx\mathcal{N}\left(\mathbf{h}_0;\frac{\mathbf{h}_t}{\sqrt{\bar{\gamma}_t}},\frac{1-\bar{\gamma}_t}{\bar{\gamma}_t}\mathbf{I}\right)$, by reparameterized sampling from this distribution, $\mathbf{h}_0$ can be equivalently written as
\begin{equation}\label{40}
	\mathbf{h}_0=
	\frac{1}{\sqrt{\bar{\gamma}_t}}
	\left(
	\mathbf{h}_t+\sqrt{1-\bar{\gamma}_t}\mathbf{u}
	\right),
	\quad
	\mathbf{u}\sim\mathcal{N}(\mathbf{0},\mathbf{I}).
\end{equation}

Substituting \eqref{40} into \eqref{eq:signal} yields
\begin{equation}
	\mathbf{y}
	=
	\frac{1}{\sqrt{\bar{\gamma}_t}}\mathbf{B}\mathbf{h}_t
	+
	\frac{\sqrt{1-\bar{\gamma}_t}}{\sqrt{\bar{\gamma}_t}}\mathbf{B}\mathbf{u}
	+\mathbf{n},
\end{equation}
from which the conditional likelihood can be approximated as
\begin{equation}
	p_t(\mathbf{y}|\mathbf{h}_t)
	=
	\mathcal{N}
	\left(
	\mathbf{y};
	\frac{1}{\sqrt{\bar{\gamma}_t}}\mathbf{B}\mathbf{h}_t,
	\frac{1-\bar{\gamma}_t}{\bar{\gamma}_t}\mathbf{B}\mathbf{B}^{T}
	+\sigma_n^2\mathbf{I}
	\right).
\end{equation}

Accordingly, the likelihood score is given by
\begin{equation}\label{eq:Likelihood}
	\begin{aligned}
		\nabla_{\mathbf{h}_t}\log p_t(\mathbf{y}|\mathbf{h}_t)
		=&
		\frac{1}{\sqrt{\bar{\gamma}_t}}
		\mathbf{B}^{T}
		\left(
		\frac{1-\bar{\gamma}_t}{\bar{\gamma}_t}\mathbf{B}\mathbf{B}^{T}
		+\sigma_n^2\mathbf{I}
		\right)^{-1}\cdot \\&
		\left(
		\mathbf{y}-\frac{1}{\sqrt{\bar{\gamma}_t}}\mathbf{B}\mathbf{h}_t
		\right).
	\end{aligned}
\end{equation}

With both the prior and likelihood scores available, channel recovery can be performed by posterior-guided iterations. Following the deterministic sampling paradigm in diffusion model \cite{song2021denoising}, the update rule is given by
\begin{equation}\label{eq:recover_new}
	\mathbf{h}_{t-1}
	=
	\frac{1}{\sqrt{\gamma_t}}
	\left(
	\mathbf{h}_t+(1-\gamma_t)\nabla_{\mathbf{h}_t}\log p_t(\mathbf{h}_t|\mathbf{y})
	\right),
	\quad t=T,\ldots,1.
\end{equation}

The proposed diffusion model-based channel estimation procedure is summarized in Algorithm~\ref{alg:DM_CE}.
\begin{algorithm}[t]
	\caption{Generative Diffusion Model-Based Algorithm for XL-MIMO Channel Estimation}
	\label{alg:DM_CE}
	\begin{algorithmic}[1]
		
		\Statex \textbf{Input:} Received signal $\mathbf{y}$, measurement matrix $\mathbf{B}$, pre-trained denoising network $\boldsymbol{\epsilon}_{\boldsymbol{\theta}}$, and noise schedule $\{\beta_t\}_{t=1}^{T}$.
		\Statex \textbf{Initialize:} $\mathbf{h}_{T}\sim\mathcal{N}(\mathbf{0},\mathbf{I})$.
		
		\For{$t=T,T-1,\ldots,1$}
		\State Compute prior score $\nabla_{\mathbf{h}_t}\log p_t(\mathbf{h}_t)$ using \eqref{eq:priori}.
		\State Compute likelihood score $\nabla_{\mathbf{h}_t}\log p_t(\mathbf{y}|\mathbf{h}_t)$ using \eqref{eq:Likelihood}.
		\State Compute posterior score $\nabla_{\mathbf{h}_t}\log p_t(\mathbf{h}_t|\mathbf{y})$ using \eqref{eq:baye}.
		\State Update $\mathbf{h}_{t-1}$ according to \eqref{eq:recover_new}.
		\EndFor
		
		\Statex \textbf{Output:} Estimated channel $\hat{\mathbf{h}}=\mathbf{h}_{0}$.
		
	\end{algorithmic}
\end{algorithm}

\subsection{Denoising Network Based on a Multi-Scale Attention Mechanism}\label{sec:network}

To better capture the spatial-frequency non-uniformity of near-field wideband channel, a denoising network with a multi-scale attention mechanism is developed for noise prediction in the diffusion model. The network takes the noisy channel sample $\mathbf{h}_t$ and the diffusion step $t$ as inputs, and treats the real and imaginary parts of the channel as two separate convolutional features. Its main purpose is to enhance multi-scale feature extraction by incorporating feature and spatial attention mechanisms.

Let the input noisy channel tensor be denoted by $\mathbf{X}_t\in\mathbb{R}^{C_{\mathrm{in}}\times N\times M}$, where $C_{\mathrm{in}}$ is the dimension of input features. The feature embedding layer $f_{FE}(\cdot)$ first maps the input into a $C$-dimensional latent space through a $3\times3$ convolution, yielding
\begin{equation}
	\mathbf{X}^0=f_{FE}(\mathbf{X}_t)\in\mathbb{R}^{C\times N\times M}.
\end{equation}

\begin{figure*}[!t]
	\centering
	\includegraphics[width=1\textwidth]{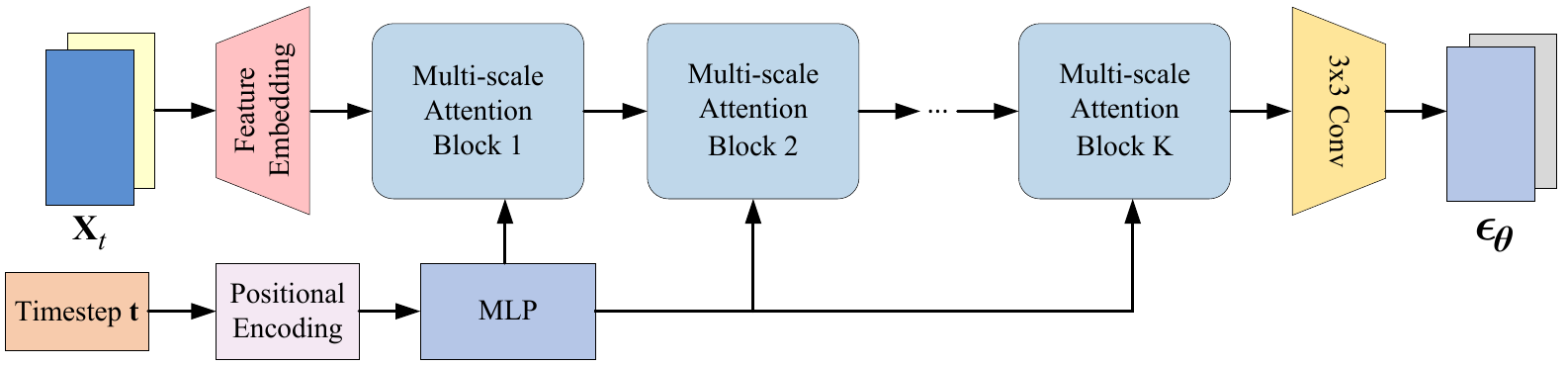}
	\caption{Architecture of the proposed denoising network with a multi-scale attention mechanism.}	
	\label{fig:F3_network1}
\end{figure*}

Subsequently, the network stacks $K$ multi-scale attention blocks to progressively extract and refine the feature representation. The output of the $k$th block is given by
\begin{equation}
	\mathbf{X}^{k}=f_{MA}^{k}(\mathbf{X}^{k-1},\mathbf{e}_t),
\end{equation}
where $\mathbf{e}_t$ is the temporal embedding vector generated at diffusion time step $t$. The extracted features are finally projected back to the original input dimension through a $3\times3$ convolution to generate the predicted noise, as shown in Fig.~\ref{fig:F3_network1}. This design enables effective noise prediction while maintaining dimensional consistency between the input and output.

\begin{figure*}[!t]
	\centering
	\includegraphics[width=1\textwidth]{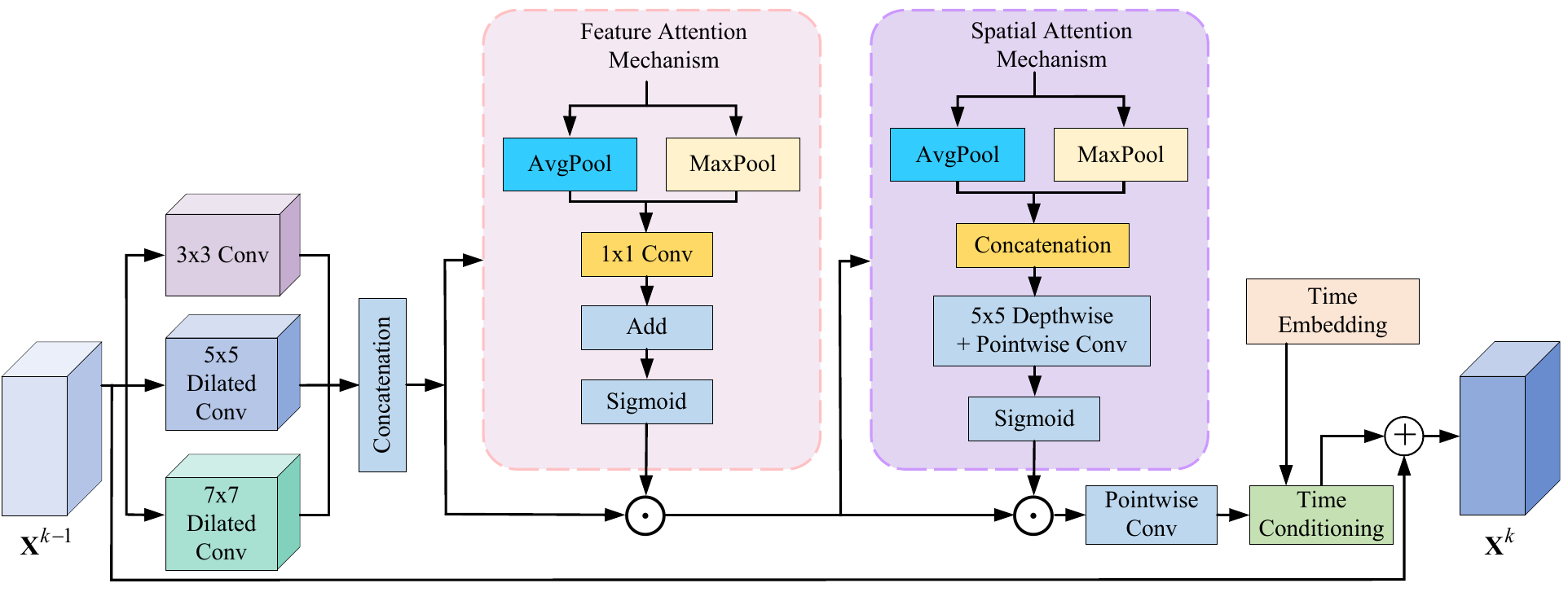}
	\caption{Detailed structure of the proposed multi-scale attention block.}	
	\label{fig:F3_network2}
\end{figure*}

Next, we describe the structure of the multi-scale attention block. As shown in Fig.~\ref{fig:F3_network2}, each block maps an input tensor in $\mathbb{R}^{C\times N\times M}$ to an output tensor of the same size. To extract multi-scale features, the input feature $\mathbf{X}^{k-1}$ is first normalized, and then processed by three parallel branches consisting of a $3\times3$ standard convolution, a $5\times5$ dilated convolution, and a $7\times7$ dilated convolution, respectively. The resulting features are concatenated along the feature dimension to form the multi-scale representation $\bar{\mathbf{X}}^{k}$, which has size $(d_1+d_2+d_3)\times N\times M$. In this way, features with different receptive fields are aggregated within a single block.

Based on $\bar{\mathbf{X}}^{k}$, a feature attention mechanism is further introduced to adaptively model feature-wise importance. Specifically, average pooling and max pooling are applied to $\bar{\mathbf{X}}^{k}$ over the spatial dimensions $(N,M)$ to generate two feature descriptors. These descriptors are passed through a shared $1\times1$ convolution, summed, and activated by a Sigmoid function to produce feature weights of size $(d_1+d_2+d_3)\times1\times1$. The resulting weights are then multiplied element-wise with $\bar{\mathbf{X}}^{k}$ to obtain the enhanced feature $\bar{\mathbf{X}}_1^{k}$, which has size $(d_1+d_2+d_3)\times N\times M$. This design enables the network to emphasize feature that are more relevant to the denoising task, thereby improving the representation capability of the multi-scale features.

To further capture the non-uniform spatial-frequency structure, a spatial attention mechanism is applied to $\bar{\mathbf{X}}_1^{k}$. Specifically, average pooling and max pooling are first performed along the feature dimension, and the resulting feature maps are concatenated and fed into a mapping layer composed of a $5\times5$ depthwise convolution and a pointwise convolution. After Sigmoid activation, a weight matrix of size $1\times N\times M$ is obtained. This weight map is multiplied element-wise with $\bar{\mathbf{X}}_1^{k}$ to generate the spatially enhanced feature $\bar{\mathbf{X}}_2^{k}$, which has size $(d_1+d_2+d_3)\times N\times M$. In this way, the network can adaptively emphasize informative regions in the spatial-frequency domain while preserving multi-scale information. Finally, a pointwise convolution is used to reduce the dimension of features in $\bar{\mathbf{X}}_2^{k}$ to $C$, yielding $\bar{\mathbf{X}}_3^{k}$.

To incorporate temporal information from the diffusion process, the diffusion step $t$ is first encoded by sinusoidal positional encoding and then mapped to a temporal embedding vector $\mathbf{e}_t$ through a multilayer perceptron. The embedding is further decomposed into a scaling term $\mathbf{s}_t$ and a bias term $\mathbf{b}_t$, which are used to modulate the feature as
\begin{equation}
	f_{time}(\bar{\mathbf{X}}_3^{k}, \mathbf{e}_t)
	=
	\bar{\mathbf{X}}_3^{k}
	+
	\mathbf{s}_t \odot \bar{\mathbf{X}}_3^{k}
	+
	\mathbf{b}_t,
\end{equation}
where $\odot$ denotes element-wise multiplication. This modulation allows the feature response to adapt to the noise level at each diffusion step.

Finally, the time-modulated feature is added to the block input through a residual connection, yielding the output of the $k$th multi-scale attention block
\begin{equation}
	\mathbf{X}^{k}
	=
	\mathbf{X}^{k-1}
	+
	f_{time}(\bar{\mathbf{X}}_3^{k}, \mathbf{e}_t).
\end{equation}
This residual design facilitates stable training and effective feature fusion across layers. Overall, by combining multi-scale attention with temporal modulation, the proposed denoising network is able to better capture the complex spatial-frequency structure of near-field wideband channel, thus providing an effective network architecture for diffusion model-based channel estimation.

\subsection{Complexity Analysis}
This subsection analyzes the computational complexity of the proposed network. Let $C$ denote the dimension of hidden features, $N\times M$ the size of the spatial-frequency feature map, and $K$ the number of stacked multi-scale attention blocks. For simplicity, let $d_1=d_2=d_3=C$. The feature embedding layer and the output layer both use standard $3\times3$ convolutions, with computational complexities of $\mathcal{O}(3^2NMC_{\mathrm{in}}C)$ and $\mathcal{O}(3^2NMC_{\mathrm{out}}C)$, respectively. The computational cost of each multi-scale attention block mainly comes from four components: multi-scale convolutions, feature attention, spatial attention, and feature compression. In the multi-scale convolution module, the input features are processed by three parallel branches with kernel sizes $3\times3$, $5\times5$, and $7\times7$, respectively, resulting in a complexity of $\mathcal{O}(3^2NMCd_1+5^2NMCd_2+7^2NMCd_3)$. In the feature attention module, two feature descriptors are obtained by global pooling, and a shared $1\times1$ convolution is used to generate feature weights, yielding a complexity of approximately $\mathcal{O}(2(d_1+d_2+d_3)^2)$. This shows that the cost of feature attention mainly depends on the feature dimension and is independent of the spatial size of the feature map. In the spatial attention module, the pooled feature maps are processed by a $5\times5$ depthwise convolution and a $1\times1$ pointwise convolution to generate a spatial weight matrix, with complexity $\mathcal{O}(2\cdot5^2NM+2NM)$. Finally, feature compression is implemented by a $1\times1$ pointwise convolution, with complexity $\mathcal{O}((d_1+d_2+d_3)CNM)$. Therefore, by neglecting lower-order terms, the overall computational complexity of the network can be approximated as $\mathcal{O}(9NMC_{\mathrm{in}}C + K(86NMC^2 + 52NM))$.

After analyzing the computational complexity, we next examine the parameter scale of the network. For the feature embedding layer and the output layer, the number of parameters is $9C_{\mathrm{in}}C$ and $9CC_{\mathrm{out}}$, respectively. The multi-scale convolutional module consists of three parallel convolutional branches, with a total of $3^2Cd_1+5^2Cd_2+7^2Cd_3$ parameters. The feature attention module has ${(d_1+d_2+d_3)}^2$ parameters. The spatial attention module consists of a $5\times5$ depthwise convolution and a $1\times1$ pointwise convolution. Since the input consists of the concatenation of two spatial descriptor matrices, its parameter count is $2\cdot5^{2}+2$. Additionally, the concatenated $3C$ features must be compressed back to $C$ dimensions via a $1\times1$ pointwise convolution, with a parameter count of $(d_1+d_2+d_3)\cdot C$. Therefore, after neglecting lower-order terms, the total number of parameters can be approximated as $9C_{\mathrm{in}}C + 95KC^2$.

Overall, the parameter size of the proposed network is mainly determined by the number of hidden features $C$ and the number of blocks $K$, whereas its computational complexity depends jointly on the feature map size $N\times M$, the feature dimension $C$, and the network depth $K$. Notably, although multi-scale convolutions are adopted to enhance feature extraction, the overall complexity grows only linearly with the feature map size. This property makes the proposed network well suited for high-dimensional channel estimation in XL-MIMO systems, where computational efficiency is of particular importance.

\section{Simulation Results and Analysis}
This section evaluates the performance of the proposed generative diffusion-based method for near-field wideband channel estimation through numerical simulations.

\subsection{Simulation Settings}
In the simulations, the base station is equipped with $N=128$ antennas, the carrier frequency is set to $60$ GHz, and the number of subcarriers is $M=64$. The angle of arrival is uniformly distributed over $\left[-\frac{\pi}{3},\frac{\pi}{3}\right]$, and the user distance is randomly selected from $[5,40]$ m. During training, the number of diffusion steps is set to $T=100$, and a linear noise schedule is adopted \cite{pmlr-v258-fesl25a}. In the denoising network, the number of features is $C_{\mathrm{in}}=C_{\mathrm{out}}=2$, the hidden feature dimension is $C=32$, and the number of multi-scale attention blocks is $K=3$. The network is trained for $100$ epochs with a batch size of $128$ using the Adam optimizer and a fixed learning rate of $5\times10^{-4}$. To ensure a fair evaluation, the training, validation, and test sets are generated independently. Specifically, $100000$ channel samples are used for training to learn the channel prior, $2000$ samples are used for validation to tune hyper parameters, and another $2000$ independently generated channel realizations are used for testing to compare the channel estimation performance of different algorithms.

This paper uses normalized mean square error (NMSE) as a performance evaluation metric, which is defined as
\begin{equation}
	\mathrm{NMSE}=\mathbb{E}\left[\frac{\|\hat{\mathbf{h}}-\mathbf{h}\|_2^2}{\|\mathbf{h}\|_2^2}\right].
\end{equation}

To comprehensively validate the effectiveness of the proposed algorithm, this paper compares it with the following representative methods:
\begin{itemize}
	\item LMMSE algorithm: This method constructs a linear estimator using the second-order channel statistics and is one of the classic benchmark methods.
	
	\item The polar-domain simultaneous orthogonal matching pursuit (PSOMP) algorithm\cite{Cui2022_Channel,11184560}: This method is based on the prior assumption of a shared sparse support set across multiple subcarriers. It formulates channel estimation as a multi-measurement-vector recovery problem under a polar-domain dictionary and utilizes the joint sparse structure of multiple subcarriers to achieve channel reconstruction. 
	
	\item The polar-domain orthogonal matching pursuit (POMP) algorithm\cite{7458188,11146467}: This method is based on polar-domain sparse representation and performs channel estimation independently at a single frequency point. It can be regarded as the single-carrier case of PSOMP.
	\item The BSPD algorithm\cite{9444239}: This method achieves wideband channel recovery in the far field by establishing a mapping relationship between physical angles and beam split patterns.
	\item The BPD algorithm\cite{10713259,cui2023near}: This method constructs an angle-distance bilinear pattern in the polar domain and recovers each near-field channel path by accumulating the largest polar-domain power from the entire bandwidth.

	\item Deep CNN algorithm\cite{8752012,11296898}: This method employs a $6$-layer $3\times3$ convolutional neural network to directly establish a mapping relationship between the received signal and the channel estimate through end-to-end training.	
\end{itemize}

\subsection{Performance Evaluation}

\begin{figure}[!t]
	\centering
	\includegraphics[width=3.5in]{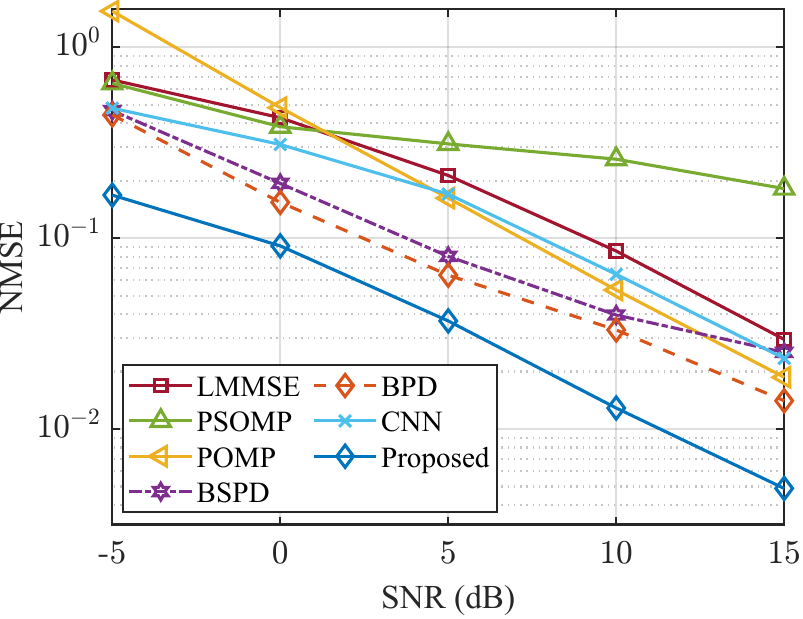}
	\caption{NMSE performance comparison of different channel estimation algorithms versus SNR for $L=4$.}
	\label{fig:F3_5}
\end{figure}

Fig. \ref{fig:F3_5} illustrates the NMSE performance of different channel estimation algorithms under different SNR conditions with the number of paths set to $L=4$. As the SNR increases, the estimation error of all methods decreases, although clear performance differences remain. In general, the LMMSE, PSOMP, and CNN methods are less effective in characterizing the complex structure of near-field wideband channels, and therefore achieve lower estimation accuracy. By contrast, the proposed diffusion-based method attains the best NMSE performance over the entire SNR range, with a particularly noticeable advantage in the low-SNR regime.
\begin{figure}[!t]
	\centering
	\includegraphics[width=3.5in]{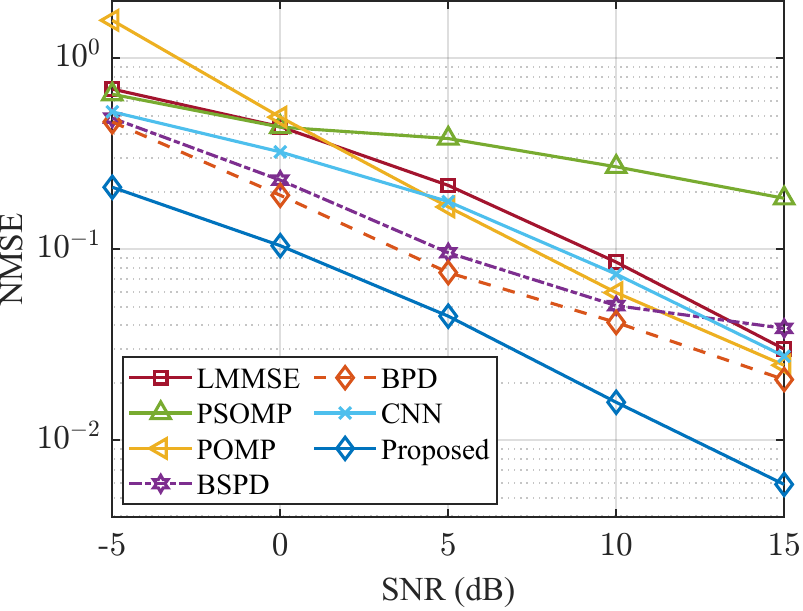}
	\caption{NMSE performance comparison of different channel estimation algorithms versus SNR for $L=6$.}	
	\label{fig:F3_6}
\end{figure} 

Specifically, the LMMSE method relies on a linear estimation framework and second-order channel statistics, which limits its ability to exploit the structural characteristics of near-field wideband channel. In addition, in practical simulations, the LMMSE estimator is typically constructed from sample covariance matrices, which may not accurately reflect the true channel statistics and thus further degrades performance. The PSOMP algorithm assumes a common sparse support across all subcarriers. However, in near-field wideband scenarios, this assumption is weakened by beam split and spatial non-stationarity, leading to performance degradation. By estimating each subcarrier independently, POMP avoids the support mismatch observed in joint multi-subcarrier recovery and therefore performs better than PSOMP. Nevertheless, POMP relies heavily on large polar codebook, resulting in considerable computational overhead. CNN method treat the channel as a two-dimensional image and perform end-to-end recovery, but their fixed receptive fields limit their ability to capture the non-uniform spatial-frequency structure of near-field wideband channel, and their performance therefore remains inferior to that of the proposed method. In contrast, the proposed method learns an implicit generative prior of the channel distribution through a diffusion model and combines it with a multi-scale attention mechanism to model spatial-frequency features at different receptive fields. As a result, it achieves higher estimation accuracy and stronger robustness, especially in the low-SNR region.

To further evaluate the channel estimation performance of different algorithms under different numbers of propagation paths, Fig.~\ref{fig:F3_6} presents the NMSE performance of various algorithms when $L=6$. It can be observed that, as the number of paths increases, the channel structure becomes more complicated, and the channel estimation task becomes correspondingly more challenging, which degrades the NMSE performance of all algorithms. Nevertheless, even when the number of paths increases, the proposed method still maintains superior estimation performance. This result indicates that the channel prior learned by the diffusion model is relatively insensitive to noise interference in complex scenarios.
\begin{figure}[!t]
	\centering
	\includegraphics[width=3.5in]{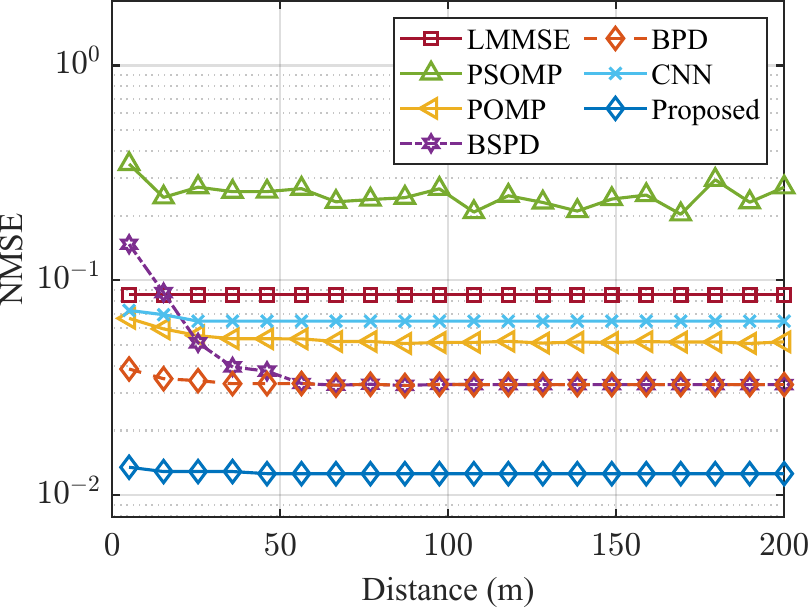}
	\caption{NMSE performance comparison of different channel estimation algorithms versus propagation distance.}	
	\label{fig:F3_7}
\end{figure}

Fig. \ref{fig:F3_7} illustrates the NMSE performance of various algorithms at different propagation distances in order to examine the impact of near-field propagation on channel estimation. In the simulations, the propagation distance is gradually increased from $5$ m to $200$ m, covering both near-field and far-field regions. The SNR is fixed at $10$ dB. As shown in Fig.~\ref{fig:F3_7}, the NMSE of the BSPD algorithm increases significantly as the user moves closer to the base station. This is because BSPD relies on an angle-domain DFT codebook for sparse channel recovery and neglects the spherical-wave effect in near-field propagation. As a result, its performance degrades rapidly once the user enters the near-field region. By contrast, PSOMP, POMP, and BPD are all based on spherical-wave modeling and therefore exhibit a certain degree of robustness over different propagation distances. Among them, although POMP and PSOMP maintain relatively stable estimation trends as the distance varies, their NMSE performance remains suboptimal because they do not effectively exploit the frequency-dependent structure induced by beam split in wideband systems. The BPD algorithm accounts for the angle-distance bilinear pattern of near-field wideband channel and therefore achieves significantly better performance than POMP and PSOMP. The proposed method outperforms all baseline algorithms over the entire distance range and demonstrates strong channel estimation capability under both near-field and far-field conditions. It can also be observed that, when the user is located close to the base station, methods based on polar codebook exhibit different degrees of performance degradation. This phenomenon is closely related to the limitations of near-field codebook design. Since it is difficult for codebook vectors to maintain good orthogonality in the distance domain, the performance of sparse recovery is inevitably affected. In contrast, the proposed method learns channel priors in a data-driven manner and employs a multi-scale attention mechanism to extract channel features under different receptive fields, thereby showing stronger robustness across a wide range of propagation distances.

\begin{figure}[!t]
	\centering
	\includegraphics[width=3.5in]{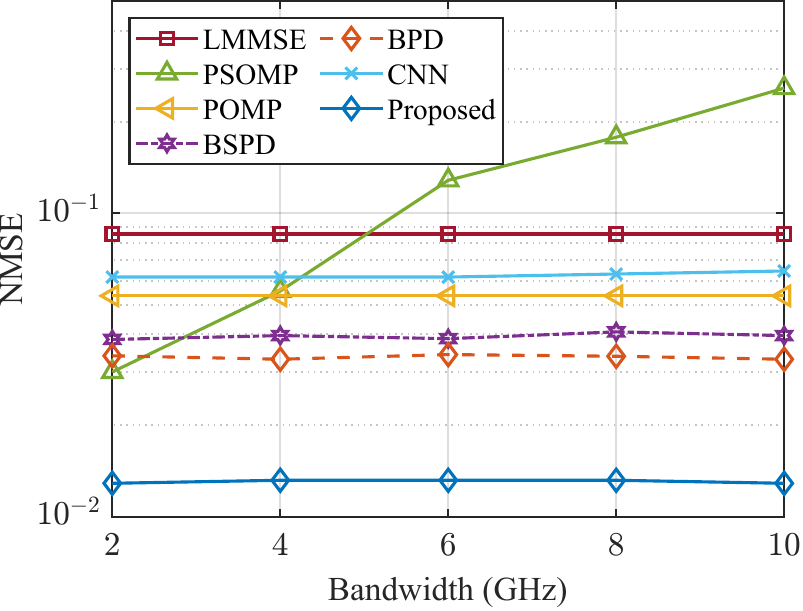}
	\caption{NMSE performance comparison of different channel estimation algorithms versus bandwidth.}	
	\label{fig:F3_8}
\end{figure}
Fig. \ref{fig:F3_8} shows the NMSE performance of various algorithms under different system bandwidths, where the bandwidth increases from $2$ GHz to $10$ GHz while the SNR is fixed at $10$ dB. The results show that, as the system bandwidth increases, the NMSE of the PSOMP algorithm rises significantly. This is because PSOMP relies on the assumption that multiple subcarriers share a common sparse support set. However, under wideband conditions, the beam split effect becomes more pronounced, and the support structures across different subcarriers are no longer consistent, which leads to progressive performance degradation. By contrast, the LMMSE algorithm does not explicitly exploit the structural characteristics of wideband channels, and its performance therefore remains relatively stable. The POMP method estimates each subcarrier independently and does not rely on the common-support assumption. As a result, it maintains a relatively stable trend as the bandwidth varies, although its overall estimation accuracy remains limited. For the BSPD algorithm, although it can describe beam split patterns in the angular domain, it does not further account for frequency-dependent variations in the range domain, and its performance gain is therefore limited in near-field wideband scenarios. Although the BPD algorithm exploits the frequency-dependent sparse structure induced by near-field beam split, its estimation performance is still constrained by the resolution of the near-field polar codebook, which limits further improvement. Overall, the proposed method consistently achieves the best performance across different bandwidths. This indicates that the designed multi-scale attention mechanism can effectively learn channel distribution features under various bandwidth conditions and adapt well to the complex structural variations caused by spatial non-stationarity and beam split in wideband scenarios.

\section{Conclusion}
This paper investigated the problem of near-field wideband channel estimation. From the perspective of channel correlation, we first analyzed the inter-antenna and inter-subcarrier correlation characteristics of near-field wideband channel. The analysis showed that such channels exhibit key properties including spherical-wave propagation, spatial non-stationarity, and beam split, which further lead to highly non-uniform structures in both the spatial and frequency domains. These characteristics make accurate modeling difficult for conventional CS methods based on hand-crafted priors, as well as for feedforward neural networks relying on simple regression mappings. To address this issue, a generative diffusion method-based for near-field wideband channel estimation is proposed. The proposed method learns an implicit prior of the channel distribution through a diffusion model and performs channel recovery within a Bayesian posterior inference framework. In addition, a denoising network with a multi-scale attention mechanism is designed to improve the modeling of complex non-uniform channel structures. Simulation results showed that the proposed method consistently outperforms existing baseline algorithms under different SNR, propagation distances, and bandwidth settings, which verifies its effectiveness for near-field wideband channel estimation.

%\bibliographystyle{IEEEtran}
%\bibliography{reference}

\newpage

\vfill

\end{document}